\documentclass[english,prd,onecolumn,superscriptaddress,nofootinbib,preprintnumbers]{revtex4}
\usepackage[latin1]{inputenc}
\usepackage{graphicx}
\usepackage{amsmath,amsthm,amssymb}
\usepackage{bm}

\makeatletter

\usepackage{amsfonts}
\usepackage{dcolumn}
\usepackage{hyperref}

\def\be{\begin{equation}}
\def\ee{\end{equation}}
\def\ba{\begin{eqnarray}}
\def\ea{\end{eqnarray}}
\def\bs{\begin{subequations}}
\def\es{\end{subequations}}

\newcommand{\rd}{{\rm d}}

\usepackage{color}

\makeatother

\usepackage{babel}
\makeatother
\begin{document}

\title{Density perturbations in general modified gravitational theories}

\author{Antonio De Felice}
\affiliation{Department of Physics, Faculty of Science, Tokyo University of Science, 
1-3, Kagurazaka, Shinjuku-ku, Tokyo 162-8601, Japan}

\author{Shinji Mukohyama}
\affiliation{Institute for the Physics and Mathematics of the Universe (IPMU), 
The University of Tokyo, Chiba 277-8582, Japan}

\author{Shinji Tsujikawa}
\affiliation{Department of Physics, Faculty of Science, Tokyo University of Science, 
1-3, Kagurazaka, Shinjuku-ku, Tokyo 162-8601, Japan}

\begin{abstract}

We derive the equations of linear cosmological perturbations 
for the general Lagrangian density $f (R,\phi, X)/2+{\cal L}_c$, 
where $R$ is a Ricci scalar, $\phi$ is a scalar field, and 
$X=-\partial^{\mu} \phi \partial_{\mu} \phi/2$ is 
a field kinetic energy.
We take into account a nonlinear self-interaction term   
${\cal L}_c=\xi (\phi) \square \phi 
(\partial^{\mu} \phi \partial_{\mu} \phi)$ 
recently studied in the context of ``Galileon'' 
cosmology, which keeps the field equations at second order.
Taking into account a scalar-field 
mass explicitly, the equations of matter density perturbations and 
gravitational potentials are obtained under a quasi-static approximation 
on sub-horizon scales.
We also derive conditions for the avoidance 
of ghosts and Laplacian instabilities associated with propagation 
speeds. Our analysis includes most of modified gravity 
models of dark energy proposed in literature and thus
it is convenient to test the viability of such models from 
both theoretical and observational points of view.

\end{abstract}

\date{\today}

\maketitle

\section{Introduction}

The constantly accumulating observational 
data \cite{SNIa,CMB,LSS} have continuously confirmed that the Universe 
entered the phase of accelerated expansion after the matter-dominated
epoch. This has motivated the idea that the gravitational law in
General Relativity (GR) may be modified at cosmological distances 
to be responsible for the cosmic acceleration 
(see Refs.~\cite{review} for review).
Many dark energy models based on the modification of 
gravity were proposed--including $f(R)$ gravity \cite{fR} and 
generalizations \cite{fRG}, scalar-tensor theory \cite{stensor}, 
ghost condensation \cite{ghostcon,PiaTsuji}, 
the Dvali-Gabadadze-Porati (DGP) braneworld model \cite{DGP}, 
and the Galileon model \cite{Galileon}.

The modified gravity models of dark energy need to be constructed
in such a way that Newton gravity is recovered at short distances 
for the consistency with solar-system experiments, 
while the deviation from GR can be allowed at large distances. 
On cosmological scales the effective mass of a scalar-field degree 
of freedom is required to be very small (of the order of the 
present Hubble constant) for realizing the cosmic acceleration today.
Such a small mass generally leads to a long range fifth force
incompatible with local gravity experiments \cite{Carroll}, 
unless some mechanism screens the scalar-field interaction 
with standard matter.

There are two known mechanisms that allow a decoupling of the 
scalar field in local regions. The first is the so-called chameleon 
mechanism \cite{chame} in which the field mass is different depending 
on the matter density in the surrounding environment.
If the field is sufficiently heavy in the regions of high density, 
a spherically symmetric body can have a ``thin-shell''
around its surface such that the effective coupling between 
the field and matter is suppressed outside the body.
In $f(R)$ gravity and scalar-tensor theory, the chameleon 
mechanism can be at work to avoid the propagation of the fifth force
even if the bare scalar-field coupling with matter is of the order 
of unity \cite{Navarro,Teg,Hu07,AT08,CapoTsuji,Brax}. 
In fact, a number of authors proposed viable dark energy models 
based on $f(R)$ and scalar-tensor theories from the requirement that 
the field mass is sufficiently large in the regions of high 
density \cite{Hu07,Star07,Appleby,TsujifR,LinderfR,TUMTY}
(see also Refs.~\cite{AGPT,LiBarrow,AT08}).

The second mechanism is the so-called Vainshtein mechanism \cite{Vainshtein}
in which nonlinear effects of scalar-field self interactions 
lead to the recovery of GR at small distances.
In massive gravity where a consistent 
massive graviton is determined by Pauli-Fierz theory \cite{Pauli}, 
the linearization close to a point-like mass source breaks down 
inside the so-called Vainshtein radius $r_*$. One needs to take into 
account a nonlinear self-interaction for the radius smaller 
than $r_*$. In the DGP braneworld model \cite{DGP} 
the Vainshtein mechanism is also at work through a  
field self-interaction of the form $(r_c^2/m_{\rm pl})\,\square 
\phi (\partial^{\mu} \phi \partial_{\mu} \phi)$, 
where $\phi$ represents a brane-bending mode, $r_c$ is a 
cross-over scale of the order of the Hubble radius $H_0^{-1}$ 
today, and $m_{\rm pl}$ is 
the Planck mass \cite{DGPVain1,DGPVain2,DGPVain3}.
Nonlinear effects lead to the decoupling of the field 
$\phi$ from matter in the region where the energy 
density $\rho$ is much larger than $r_c^{-2}m_{\rm pl}^2$.

Although the Vainshtein effect screens the fifth force mediated 
by the brane-bending mode, the self-accelerating cosmological 
solution in the DGP model is plagued by 
the appearance of a ghost mode \cite{DGPVain3,Nicolis04,Gorbunov}.
In order to realize self-accelerating solutions without ghosts, 
the equations of motion should be kept at second order in derivatives.
This can be addressed by considering higher 
derivative scalar-field interactions that respect a constant gradient-shift in the Minkowski 
space-time, i.e. $\partial_{\mu} \phi \to \partial_{\mu} \phi+b_{\mu}$ \cite{Nicolis}.
In addition to the term proportional to $\square \phi (\partial^{\mu} \phi \partial_{\mu} \phi)$, 
Nicolis {\it et al.} \cite{Nicolis} derived other higher derivative interactions that 
keep the equations of motion at second order. 
This was extended to a more generalized covariant 
Galileon nonminimally coupled to the scalar curvature \cite{Deffayet}.

In the presence of a nonlinear self-interaction of the form 
 $\xi (\phi) \square \phi (\partial^{\mu} \phi \partial_{\mu} \phi)$, 
where $\xi (\phi)$ is a function of a scalar field $\phi$, 
Silva and Koyama \cite{Silva} showed that 
it is possible to avoid the appearance of a ghost mode in the 4-dimensional 
Brans-Dicke theory \cite{BDicke}.
Although such a term does not satisfy the Galilean invariance in the 
Friedmann-Lema\^{i}tre-Robertson-Walker (FLRW) cosmological background, 
the equations of motion remain at second order.
For the choice $\xi (\phi) \propto \phi^{-2}$ there exists a de Sitter solution 
responsible for the late-time cosmic acceleration, while the Vainshtein 
mechanism allows to recover the General Relativistic behavior 
in the asymptotic past \cite{Silva}.
Recently this was extended to more general modified gravitational 
theories in which the late-time de Sitter solution is realized by the field 
kinetic energy \cite{DeFelice10}.

In order to confront dark energy models with observations of 
large-scale structure, cosmic microwave background (CMB), 
and weak lensing, it is important to study the evolution of cosmological 
perturbations \cite{obsermo}.
If the gravity is modified from GR, the effective gravitational coupling 
characterizing the growth rate of matter perturbations is 
subject to change.
Hence, in general, modified gravity models leave distinct observational 
signatures compared to models based on GR.
In this respect the evolution of matter  
perturbations has been extensively studied for a number of 
modified gravity models--including $f(R)$ gravity \cite{fRper,Star07,Hu07,Tsujikawa07,TsujifR}, 
scalar-tensor theory \cite{Boi,Uzan,Verde,Gannouji,TUMTY,Koyama10}, the DGP model \cite{DGPper}, 
and the Galileon model \cite{Silva,Chow,Kobayashi,GanSami} 
(see also Refs.~\cite{Nesseris}).
In Ref.~\cite{Tsujikawa07} the equation of matter perturbations 
was derived for the general Lagrangian $f_1(R,\phi)/2+f_2(\phi,X)/2$,
under the assumption that the mass of the field $\phi$ is as light 
as the Hubble constant $H_0$ today. This assumption can be 
justified for the field potential having a light mass during most of
the cosmological epochs. In scalar-tensor models with a 
coupling of the order of unity between the field $\phi$ and matter, 
the potential needs to be designed
to have a large mass in the regions of high density so that the chameleon 
mechanism screens the fifth force \cite{TUMTY}.
Cosmologically the transition from 
the ``massive'' regime to the ``massless'' regime can occur during the 
matter era, depending on the wavelength of perturbations (including 
$f(R)$ gravity) \cite{Star07,TsujifR,TUMTY,Koyama10}.
Hence it is important to take into account such a mass term 
to estimate the epoch of transition and the resulting matter 
power spectrum.

In this paper we study the evolution of density perturbations for the  
general action (\ref{action}) given below. 
In addition to the effect of the field mass, we discuss
the effect of the nonlinear self-interaction 
${\cal L}_c=\xi (\phi) \square \phi (\partial^{\mu} \phi \partial_{\mu} \phi)$
on the dynamics of perturbations.
In other words we consider two effects for the recovery of GR 
in the past cosmological evolution--(i) the chameleon mechanism
through the field mass $M_\phi$, and (ii) the Vainshtein mechanism 
through the self-interaction term ${\cal L}_c$.
For the theories with $f=f_1 (R,\phi)+f_2 (\phi, X)$ plus 
${\cal L}_c$, we shall derive the equation of matter perturbations as well 
gravitational potentials under the quasi-static 
approximation \cite{Star98,Boi,Tsujikawa07,Koyama10} 
on sub-horizon scales.
Since our analysis is sufficiently general, this will be useful to place observational 
constraints on each model.
We also derive conditions for the avoidance of ghosts
and Laplacian instabilities, without 
recourse to any approximation.
This is required for the construction of viable dark energy 
models free from theoretical inconsistency.

\section{Perturbation equation in general modified gravitational theories}

We start with the following 4-dimensional action
\begin{equation}
\label{action}
S=\int {\rm d}^4 x \sqrt{-g} \left[
\frac{1}{2}f(R, \phi, X)+
\xi (\phi) \square \phi (\partial^{\mu} \phi 
\partial_{\mu} \phi) \right]
+\int {\rm d}^4 x {\cal L}_m (g_{\mu \nu}, \Psi_m)\,,
\end{equation}
where $g$ is the determinant of the metric $g_{\mu \nu}$,
$f$ is a function in terms of the Ricci scalar $R$, 
a scalar field $\phi$ and a kinetic term 
$X =-g^{\mu \nu}\partial_{\mu}\phi \partial_{\nu} \phi/2$, 
and $\xi (\phi)$ is a function of $\phi$.
${\cal L}_m$ is a matter Lagrangian that depends on 
the metric $g_{\mu \nu}$ and matter fields $\Psi_m$.
If $\xi (\phi)$ is constant, then this term respects the Galilean
symmetry $\partial_{\mu} \phi \to \partial_{\mu} \phi+b_{\mu}$ as well
as the shift symmetry $\phi\to\phi +c$ 
in the Minkowski space-time \cite{Nicolis,Deffayet,Chow}.
Here we consider a general function $\xi (\phi)$ in terms of $\phi$, 
as this allows the possibility to give rise to the late-time cosmic acceleration
without the appearance of a ghost \cite{Silva,Kobayashi}.

Varying the action (\ref{action}) with respect to $g_{\mu \nu}$ and $\phi$, 
we obtain the following field equations 
\begin{eqnarray}
& & FG_{\mu \nu}=\frac12 (f-RF) g_{\mu \nu}+\nabla_{\mu} \nabla_{\nu}F
-g_{\mu \nu} \square F+\frac12 f_{,X} \nabla_{\mu} \phi \nabla_{\nu}\phi+
T_{\mu \nu}^{(c)}+T_{\mu \nu}^{(m)}\,, \\
& & \nabla_{\mu} \left( f_{,X} \nabla^{\mu} \phi \right)
+f_{,\phi}=T^{(c)}\,,
\end{eqnarray}
where $F \equiv \partial f/\partial R$, $f_{,X} \equiv \partial f/\partial X$,
$f_{,\phi} \equiv \partial f/\partial \phi$, 
$T_{\mu \nu}^{(m)} \equiv -(2/\sqrt{-g}) \delta {\cal L}_m/\delta g^{\mu \nu}$ 
is the energy-momentum tensor of matter, and 
\begin{eqnarray}
& & T_{\mu \nu}^{(c)}=2 \left( \xi \partial^{\lambda} \phi
\partial_{\lambda} \phi \right)_{,}{}_{(\mu} \phi_{,\nu)}
-2\xi \square \phi (\partial_{\mu} \phi \partial_{\nu} \phi)
-\partial_{\lambda} (\xi \partial^{\rho}\phi \partial_{\rho}\phi)
\partial^{\lambda} \phi\,g_{\mu \nu}\,, \\
& & T^{(c)}=-2 \left[ \xi_{,\phi} \square \phi 
( \partial^{\mu} \phi \partial_{\mu} \phi)+
\square \left( \xi \partial^{\mu} \phi \partial_{\mu} \phi \right)
-2\nabla_{\mu} \left(\xi\,\square \phi\,\partial^{\mu} \phi \right)
\right]\,.
\end{eqnarray}

We consider the following perturbed metric about a spatially flat 
FLRW cosmological background with scalar 
metric perturbations $\Psi$ and $\Phi$
in a longitudinal gauge \cite{Bardeen}:
\begin{equation}
\rd s^2 =- (1+2\Psi)\rd t^2 
+a(t)^2 (1-2\Phi)\delta_{ij}\rd x^i \rd x^j\,,
\label{permet}
\end{equation}
where $a(t)$ is a scale factor with cosmic time $t$. 
We caution that our notations of $\Psi$ and $\Phi$ are opposite to 
those used in Ref.~\cite{Tsujikawa07}. 
We decompose the field $\phi$ into the background and inhomogeneous
parts: $\phi (t, {\bm x})=\tilde{\phi}(t)+\delta \phi (t, {\bm x})$.
We consider a perfect fluid for the energy-momentum tensors
$T_{\mu \nu}^{(m)}$, so that they can be decomposed as 
${T^{0}_{0}}^{(m)}=-(\rho+\delta \rho)$, 
${T^0_{\alpha}}^{(m)}=-(\rho+P) v_{, \alpha}$, and 
${T^{\alpha}_{\beta}}^{(m)}=(P+\delta P) \delta^{\alpha}_{\beta}$
(where $v$ is a velocity potential).
We also decompose $T_{\mu \nu}^{(c)}$ and $T^{(c)}$
into the background and perturbative parts, as 
$T_{\mu \nu}^{(c)}=\tilde{T}_{\mu \nu}^{(c)}
+\delta T_{\mu \nu}^{(c)}$ and 
$T^{(c)}=\tilde{T}^{(c)}+\delta T^{(c)}$.
In the following, when we express background quantities, we drop 
the tilde for simplicity.

In the flat FLRW background we obtain the following equations
\begin{eqnarray}
\label{be1}
& & 3FH^2=f_{,X}X+\frac12 (FR-f)-3H\dot{F}
+(6H\xi-\xi_{,\phi} \dot{\phi}) \dot{\phi}^3+\rho\,,\\
\label{be2}
& & 3FH^2+2F\dot{H}=\frac12 (FR-f)-\ddot{F}-2H\dot{F}
+\dot{\phi}^2 (2\xi \ddot{\phi}+\xi_{,\phi} \dot{\phi}^2)-P\,,\\
\label{be3}
& & f_{,X}\ddot{\phi}+\left( 3Hf_{,X}+\dot{f_{,X}} \right)
\dot{\phi}-f_{,\phi}+T^{(c)}=0\,, \\
\label{be4}
& & \dot{\rho}+3H(\rho+P)=0\,,
\end{eqnarray}
where a dot represents a derivative with respect to $t$, and 
\begin{eqnarray}
& & H=\dot{a}/a\,,\qquad 
R=6(2H^2+\dot{H})\,,\\
& & T^{(c)}=-2\dot{\phi}
\left[ \dot{\phi} (4 \xi_{,\phi} \ddot{\phi}+\xi_{,\phi \phi} \dot{\phi}^2)
-6 \xi \left\{ 2H\ddot{\phi}+(3H^2+\dot{H}) \dot{\phi} \right\}
\right]\,.
\end{eqnarray}

We also obtain the following linearized equations for 
the perturbed metric (\ref{permet}) in the Fourier 
space\footnote{We checked that $\delta {T^0_0}^{(c)}$ and $\delta T^{(c)}$ 
in Eqs.~(\ref{T00}) and (\ref{Tc}) coincide
with those given in Ref.~\cite{Hwang}. However there are some differences in 
Eqs.~(\ref{per1}) and (\ref{per2}) compared to Ref.~\cite{Hwang}.
We do not have the term $(\delta F/F){T^0_0}^{(c)}$ in Eq.~(\ref{per1})
and we have the term $2\Psi T^{(c)}$ in Eq.~(\ref{per2}).}:   
\begin{eqnarray}
\label{per1}
& &3H(\dot{\Phi}+H\Psi)+\frac{k^2}{a^2}\Phi +\frac{1}{2F}
\biggl[ -\frac12 (f_{,\phi}\delta \phi+f_{,X}\delta X )+
\frac12 \dot{\phi}^2 (f_{,X \phi} \delta \phi +f_{,XX}\delta X
+F_{,X} \delta R)
+f_{,X}\dot{\phi}\dot{\delta \phi} 
-3H\dot{\delta F} \nonumber \\
& &+\left(3H^2+3\dot{H}-\frac{k^2}{a^2}
\right) \delta F+3\dot{F} (\dot{\Phi}+H\Psi)+
(3H\dot{F}-f_{,X}\dot{\phi}^2)\Psi
-{\delta T^{0}_{0}}^{(c)}
+\delta \rho \biggr]=0\,,\\
\label{per2}
& & f_{,X} \left[ \ddot{\delta \phi}+\left( 3H+
\frac{\dot{f}_{,X}}{f_{,X}} \right) \dot{\delta \phi}
+\frac{k^2}{a^2}\delta \phi -\dot{\phi} 
(3\dot{\Phi}+\dot{\Psi}) \right]
-2f_{,\phi}\Psi+\frac{1}{a^3} (a^3 \dot{\phi}
\delta f_{,X})^{\cdot}-\delta f_{,\phi}=
-\delta T^{(c)}-2\Psi T^{(c)}
\,,\\
\label{per3}
& &\Phi=\Psi+\frac{\delta F}{F}\,,\\
\label{per4}
& & \delta \dot{\rho}+3H \left( \delta \rho
+\delta P \right)
=(\rho+P) \left( 3\dot{\Phi}-\frac{k^2}{a^2}v \right) 
\,, \\
\label{per5}
& & \frac{1}{a^3 (\rho+P)} \frac{{\rm d}}{{\rm d}t}
\left[ a^3 (\rho+P)v \right]=\Psi+\frac{\delta P}
{\rho+P}\,,
\end{eqnarray}
where $k$ is a comoving wavenumber, and
\begin{eqnarray}
\label{T00}
{\delta T^{0}_{0}}^{(c)}&=& \dot{\phi}^2
\biggl[ 3(\xi_{,\phi} \dot{\phi}-6H \xi) \dot{\delta \phi}
-2\xi \frac{k^2}{a^2}\delta \phi+\dot{\phi} 
\left\{ \xi_{,\phi} \dot{\delta \phi}+(\xi_{,\phi \phi}
\dot{\phi}-6H \xi_{,\phi}) \delta \phi \right\} \nonumber \\
& &~~~~+6 \xi \dot{\phi} (\dot{\Phi}+H\Psi)
-2\dot{\phi} (2\xi_{,\phi} \dot{\phi}-9H \xi ) \Psi \biggr]\,,
\\
\label{Tc}
\delta T^{(c)}&=&-2\, \biggl\{ 4\dot{\phi} ( \xi_{,\phi} \dot{\phi}
-3H\xi) \ddot{\delta \phi}
+[4\dot{\phi} (\xi_{,\phi \phi} \dot{\phi}^2+2\xi_{,\phi} \ddot{\phi})
-12 \xi (H\ddot{\phi}+3H^2 \dot{\phi}+\dot{H} \dot{\phi})]\dot{\delta \phi}
\nonumber \\
& & 
~~~~~~+\left[ \dot{\phi}^2 (\xi_{,\phi \phi \phi} \dot{\phi}^2+4\xi_{,\phi \phi} 
\ddot{\phi})-6\xi_{,\phi} \dot{\phi} (2H\ddot{\phi}+3H^2 \dot{\phi}
+\dot{H} \dot{\phi}) -4\xi (\ddot{\phi}+2H\dot{\phi})
\frac{k^2}{a^2}\right] \delta \phi \nonumber \\
& &~~~~~~+6\xi \dot{\phi} \biggl[ 2(\dot{\Phi}+H\Psi)(\ddot{\phi}+
3H\dot{\phi})+\dot{\phi} (\ddot{\Phi}+\dot{H}\Psi+H \dot{\Psi})
\biggr]-4\dot{\phi}^2 (\xi_{,\phi} \dot{\phi}-3H\xi) \dot{\Psi} \nonumber \\
& &~~~~~~-4\dot{\phi}^2 (\xi_{,\phi \phi} \dot{\phi}^2+4 \xi_{,\phi} \ddot{\phi})
\Psi+2\xi \dot{\phi} \left[ 9(2H \ddot{\phi}+2H^2\dot{\phi}
+\dot{H} \dot{\phi})-\dot{\phi} \frac{k^2}{a^2} \right] \Psi 
\biggr\}\,.
\end{eqnarray}
We introduce the gauge-invariant matter density perturbation $\delta_m$, as 
\begin{equation}
\delta_m \equiv \frac{\delta \rho}{\rho}
+3H(1+w)v\,,
\end{equation}
where $w \equiv P/\rho$.

\section{Quasi-static approximation for perturbations of non-relativistic matter}

The action (\ref{action}) covers a wide variety of 
dark energy theories such as $f(R)$ gravity \cite{fR}, Brans-Dicke theory \cite{BDicke}, 
scalar-tensor theory \cite{stensor}, and k-essence \cite{kes}
(ghost condensate, tachyon, Dirac-Born-Infeld models, etc).
Most of those theories take the Lagrangian of the form
\begin{equation}
\label{Lag}
f(R,\phi,X)=f_1(R,\phi)+f_2(\phi,X)\,.
\end{equation}
In this case the quantity
\begin{equation}
F=\frac{\partial f}{\partial R}=
\frac{\partial f_1(R,\phi)}{\partial R}\,,
\end{equation}
is a function of $R$ and $\phi$.
Note that the Lagrangian (\ref{Lag}) does not include theories in which 
the kinetic term $X$ couples to the Ricci scalar $R$.

In Ref.~\cite{Tsujikawa07} the perturbation equation for non-relativistic  
matter has been derived for the Lagrangian (\ref{Lag})
with $\xi (\phi)=0$, under the assumption that the effective 
mass of the field $\phi$ is smaller than 
the Hubble expansion rate $H$.
In this work we shall take into account the effective 
mass $M_{\phi}$ of the field by defining 
\begin{equation}
M_{\phi}^2 \equiv -f_{,\phi \phi}/2\,.
\end{equation}
For a minimally coupled scalar field with the Lagrangian 
$f=R/(8\pi G)+2X-2V(\phi)$, we have that  $M_{\phi}^2=V_{,\phi \phi}$.
The quantity $M_\phi$ does not in general 
correspond to the mass of the propagating canonical field. 

For the perfect fluid let us consider non-relativistic matter ($w=0$). 
Combining Eqs.~(\ref{per4}) and (\ref{per5}), the gauge-invariant 
matter perturbation $\delta_m=\delta \rho/\rho+3Hv$ satisfies
\begin{equation}
\label{delright}
\ddot{\delta}_m+2H\dot{\delta}_m+\frac{k^2}{a^2}\Psi
=3\ddot{B}+6H\dot{B}\,,
\end{equation}
where $B \equiv \Phi+Hv$.
The modes relevant to the observations of large-scale 
structure correspond to the sub-horizon perturbations
($k \gg aH$). In order to derive the equation of matter perturbations 
approximately, we use the quasi-static approximation 
under which the dominant terms in Eqs.~(\ref{per1}), (\ref{per2}), and 
(\ref{delright}) are  those including $k^2/a^2$,
$\delta \rho$ (or $\delta_m$), and 
$M_{\phi}^2$ \cite{Star98,Boi,Tsujikawa07,Koyama10}.
As long as the oscillating mode of a scalar-field degree of
freedom is suppressed relative to the mode induced by 
matter perturbations, this approximation is valid 
for sub-horizon perturbations \cite{Star07,TsujifR,TUMTY}.
First of all, the perturbation in the Ricci scalar is given by 
\begin{eqnarray}
\delta R &=& 2 \left[ \left( \frac{k^2}{a^2}-3\dot{H} \right)\Psi
-2\frac{k^2}{a^2}\Phi-3 (\ddot{\Phi}+4H\dot{\Phi}
+H\dot{\Psi}+\dot{H}\Psi+4H^2 \Psi ) \right] \nonumber \\
&\simeq& 2\frac{k^2}{a^2} (\Psi-2\Phi)\,.
\label{delRap}
\end{eqnarray}
{}From Eqs.~(\ref{per3}) and (\ref{delRap}) it follows that 
\begin{equation}
\label{delR2}
\delta R \simeq -\frac{2k^2}{a^2} \frac{\Psi+(2F_{,\phi}/F)\delta \phi}
{1+4r_1}\,,
\end{equation}
where 
\begin{equation}
r_1 \equiv \frac{k^2}{a^2} \frac{F_{,R}}{F}
\simeq \frac{k^2}{3a^2}\frac{1}{M_R^2}\,.
\end{equation}
Here $M_R^2 \equiv F/(3F_{,R})$ is the mass squared
of a scalar-field degree of freedom coming from the 
curvature term \cite{Navarro,Star07} 
(which is valid in the region $M_R^2 \gg H^2$).

For the theory (\ref{Lag}) the perturbation $\delta f$
is given by 
\begin{equation}
\delta f=F\delta R+f_{,\phi}\delta \phi+f_{2,X}\delta X\,,
\end{equation}
where $\delta X=\dot{\phi} \dot{\delta \phi}-\dot{\phi}^2 \Psi$.
Under the quasi-static approximation we have 
\begin{equation}
\delta f_{,\phi} \simeq F_{,\phi}\delta R
-2M_{\phi}^2 \delta \phi\,.
\end{equation}
{}From Eq.~(\ref{per2}) we find
\begin{equation}
\label{fieldre}
f_{,X} \frac{k^2}{a^2} \delta \phi-F_{,\phi} \delta R
+2M_{\phi}^2 \delta \phi \simeq -\delta T^{(c)}\,,
\end{equation}
where 
\begin{equation}
\delta T^{(c)} \simeq 4\xi \frac{k^2}{a^2}
\left[ 2(\ddot{\phi}+2H \dot{\phi}) \delta \phi
+\dot{\phi}^2 \Psi \right]\,.
\end{equation}
Here we have neglected the oscillating mode of the field 
perturbation $\delta \phi$.
If the field is sufficiently heavy in the regions of high density,
such an oscillating mode can be important as we go back to 
the past \cite{Star07,TsujifR,Battye}.
The initial conditions for the field perturbation need to be 
chosen so that the oscillating mode is suppressed 
relative to the matter-induced mode.

{}From Eqs.~(\ref{delR2}) and (\ref{fieldre}), it follows that 
\begin{eqnarray}
\label{delphieq}
\delta \phi &\simeq& -\frac{2F_{,\phi}+4\xi \dot{\phi}^2 (1+4r_1)}
{(1+4r_1) [ f_{,X}+2r_2+8 \xi (\ddot{\phi}+2H\dot{\phi})]
+4F_{,\phi}^2/F}\Psi\,,\\
\label{delReq}
\delta R &\simeq& -\frac{2k^2}{a^2}
\frac{f_{,X}+2r_2+8 \xi (\ddot{\phi}+2H\dot{\phi}-\dot{\phi}^2F_{,\phi}/F)}
{(1+4r_1) [ f_{,X}+2r_2+8 \xi (\ddot{\phi}+2H\dot{\phi})]
+4F_{,\phi}^2/F}\Psi\,,
\end{eqnarray}
where 
\begin{equation}
r_2 \equiv \frac{a^2}{k^2}M_\phi^2\,.
\end{equation}
{}From Eq.~(\ref{per1}) one has
\begin{equation}
\frac{k^2}{a^2}\Phi \simeq \frac{1}{2F}
\left( \frac{k^2}{a^2} \delta F-\delta \rho-2\xi \dot{\phi}^2
\frac{k^2}{a^2} \delta \phi \right)\,,
\end{equation}
which, together with Eq.~(\ref{per3}), gives 
\begin{equation}
\label{Psieq}
\frac{k^2}{a^2}\Psi \simeq -\frac{1}{2F}
\left( \frac{k^2}{a^2} \delta F+\delta \rho+2\xi \dot{\phi}^2
\frac{k^2}{a^2} \delta \phi \right)\,.
\end{equation}

Plugging Eqs.~(\ref{delphieq}) and (\ref{delReq}) into 
Eq.~(\ref{Psieq}), we obtain 
\begin{eqnarray}
\label{Psieqfi}
\frac{k^2}{a^2}\Psi &\simeq& 
-\frac{\delta \rho}{2F} \frac{(1+4r_1)(f_{,X}+2r_2)+4F_{,\phi}^2/F
+8 \xi (1+4r_1)(\ddot{\phi}+2H\dot{\phi})}{(1+3r_1)(f_{,X}+2r_2)+
3F_{,\phi}^2/F+2\xi [4(1+3r_1)(\ddot{\phi}+2H\dot{\phi})
-2F_{,\phi}\dot{\phi}^2/F-2\xi \dot{\phi}^4 (1+4r_1)/F]}
\,,\\
\label{Phieqfi}
\frac{k^2}{a^2}\Phi &\simeq& 
-\frac{\delta \rho}{2F} \frac{(1+2r_1)(f_{,X}+2r_2)+2F_{,\phi}^2/F
+4 \xi [2(1+2r_1)(\ddot{\phi}+2H \dot{\phi})-F_{,\phi}\dot{\phi}^2/F]}
{(1+3r_1)(f_{,X}+2r_2)+3F_{,\phi}^2/F+2\xi [4(1+3r_1)(\ddot{\phi}+2H\dot{\phi})
-2F_{,\phi}\dot{\phi}^2/F-2\xi \dot{\phi}^4 (1+4r_1)/F]}\,.
\end{eqnarray}
Under the quasi-static approximation on sub-horizon scales the r.h.s. of 
Eq.~(\ref{delright}) can be neglected relative to the l.h.s. of it.
Moreover the gauge-invariant matter perturbation can be 
approximated as $\delta_m \simeq \delta \rho/\rho$.
Then, from Eqs.~(\ref{delright}) and (\ref{Psieqfi}), 
we finally obtain 
\begin{equation}
\label{delmori}
\ddot{\delta}_m+2H \dot{\delta}_m
-4\pi G_{\rm eff}\,\rho \,\delta_m \simeq 0\,.
\end{equation}
Here the effective gravitational coupling is given by 
\begin{equation}
\label{Geff}
G_{\rm eff} = \frac{1}{8\pi F}\frac{1+4r_1}{1+3r_1}
\left\{ 1 + \frac{[F_{,\phi}+2(1+4r_1)\xi\dot{\phi}^2]^2}
{(1+4r_1)\mu F} \right\}\,,
\end{equation}
where 
\begin{equation}
\label{muexp}
\mu \equiv (1+3r_1)(f_{,X}+2r_2)+3F_{,\phi}^2/F
+2\xi [4(1+3r_1)(\ddot{\phi}+2H\dot{\phi})
-2F_{,\phi}\dot{\phi}^2/F-2\xi \dot{\phi}^4 (1+4r_1)/F]\,.
\end{equation}

We define the following quantity 
\begin{equation}
\zeta \equiv \Psi/\Phi\,.
\end{equation}
From Eqs.~(\ref{Psieqfi}) and (\ref{Phieqfi}) it follows that  
\begin{equation}
\label{zeta}
\zeta \simeq \frac{(1+4r_1)(f_{,X}+2r_2)+4F_{,\phi}^2/F
+8 \xi (1+4r_1)(\ddot{\phi}+2H\dot{\phi})}{(1+2r_1)(f_{,X}+2r_2)+2F_{,\phi}^2/F
+4 \xi [2(1+2r_1)(\ddot{\phi}+2H \dot{\phi})-F_{,\phi}\dot{\phi}^2/F]}\,.
\end{equation}
We also introduce the effective gravitational potential 
\begin{equation}
\Phi_{\rm eff}=(\Psi+\Phi)/2\,.
\end{equation}
This quantity characterizes the deviation of light rays, which is linked with 
the Integrated Sachs-Wolfe (ISW) effect in CMB
and weak lensing observations \cite{Kunz}. 
{}From Eqs.~(\ref{Psieqfi}) and (\ref{Phieqfi}) we have 
\begin{equation}
\label{Phieff}
\Phi_{\rm eff} \simeq -\frac{\rho\,\delta_m}{2F} \frac{a^2}{k^2} 
\frac{(1+3r_1)[f_{,X}+2r_2+8\xi (\ddot{\phi}+2H\dot{\phi})]+3F_{,\phi}^2/F
-2\xi F_{,\phi} \dot{\phi}^2/F}
{(1+3r_1)[f_{,X}+2r_2+8\xi (\ddot{\phi}+2H\dot{\phi})]+3F_{,\phi}^2/F
-4\xi [F_{,\phi} \dot{\phi}^2/F+\xi \dot{\phi}^4 (1+4r_1)/F]}\,.
\end{equation}
Note that for the theories with $\xi=0$ this reduces to 
\begin{equation}
\label{Phieff0}
\Phi_{\rm eff}\simeq -\frac{\rho\,\delta_m}{2F} \frac{a^2}{k^2}\,.
\end{equation}
The presence of the $\xi$ term leads to the modification to 
the effective gravitational potential and hence this can leave some
nontrivial signature on the CMB spectrum \cite{Silva,Kobayashi}.

\section{Evolution of perturbations in specific theories}
\label{specific}

In this section we apply the results in the previous section to a number of 
concrete modified gravitational theories.
It is convenient to write the matter perturbation 
equation (\ref{delmori}) in the form 
\begin{equation}
\delta_m''+\left( \frac12 -\frac32 w_{\rm eff} \right)\delta_m'
-(12\pi F G_{\rm eff}) \Omega_m \delta_m \simeq 0\,,
\end{equation}
where a prime represents a derivative with respect to 
$N \equiv \ln a$, and 
\begin{equation}
w_{\rm eff} \equiv -1-\frac{2\dot{H}}{3H^2}\,,\qquad
\Omega_m \equiv \frac{\rho}{3FH^2}\,.
\end{equation}
\subsection{$f(R)$ gravity with $\xi=0$}

Let us consider the theories with the Lagrangian 
\begin{equation}
f(R, \phi, X)=f_1(R)+f_2 (\phi, X)\,,
\end{equation}
which includes $f(R)$ gravity as a specific case.
When $\xi=0$, Eqs.~(\ref{Psieqfi}), (\ref{Phieqfi}), (\ref{Geff}), 
and (\ref{zeta}) give
\begin{equation}
\frac{k^2}{a^2}\Psi \simeq -\frac{\delta \rho}{2F} \frac{1+4r_1}{1+3r_1}\,,
\qquad
\frac{k^2}{a^2}\Phi \simeq -\frac{\delta \rho}{2F} \frac{1+2r_1}{1+3r_1}\,, 
\qquad
G_{\rm eff} \simeq \frac{1}{8\pi F} \frac{1+4r_1}{1+3r_1}\,,\qquad
\zeta=\frac{1+4r_1}{1+2r_1}\,,
\end{equation}
where $\Phi_{\rm eff}$ is given in Eq.~(\ref{Phieff0}).
This matches with the results obtained in Refs.~\cite{Zhang05,Tsujikawa07}
in the context of $f(R)$ gravity.
Even in the presence of a scalar field $\phi$ minimally 
coupled to gravity, the above results are the same 
as those in $f(R)$ gravity.

In the regime where the mass squared $M_R^2$ of the scalar-field 
degree of freedom (scalaron \cite{Star80}) is much larger than $k^2/a^2$, i.e.
$r_1 \ll 1$, the evolution of perturbations during the matter dominance
($a \propto t^{2/3}$, $w_{\rm eff} \simeq 0$, and 
$\Omega_m \simeq 1$) is given by $\delta_m \propto t^{2/3}$, 
$\Phi_{\rm eff} \simeq$\,constant, and $\zeta \simeq 1$.
In the regime $M_R^2 \ll k^2/a^2$, the perturbations during the 
matter era evolve as \cite{Star07,TsujifR}
\begin{equation}
\label{delmfR}
\delta_m \propto t^{(\sqrt{33}-1)/6}\,,\qquad
\Phi_{\rm eff} \propto t^{(\sqrt{33}-5)/6}\,,\qquad
\zeta \simeq 2\,.
\end{equation}
After the perturbations enter the epoch of cosmic acceleration, 
the evolution (\ref{delmfR}) is subject to change.

\subsection{Scalar-tensor theories with $\xi=0$}

Let us consider scalar-tensor theories with $\xi=0$
in which the function $f_1$ in the Lagrangian (\ref{Lag}) 
depends on both $R$ and $\phi$.
This covers the theories with the effective
two-scalar degrees of freedom, i.e.\ scalaron 
(the gravitational scalar) and the field $\phi$.
The scalar-tensor theory with $f=F(\phi)R+f_2 (\phi,X)$
has one scalar degree of freedom, but the nonlinear action 
in $R$ gives rise to another gravitational scalar degree of freedom.
{}From Eqs.~(\ref{Psieqfi}), (\ref{Phieqfi}), (\ref{Geff}), 
and (\ref{zeta}), it follows that 
\begin{eqnarray}
& & \frac{k^2}{a^2}\Psi \simeq 
-\frac{\delta \rho}{2F} \frac{(1+4r_1)(f_{,X}+2r_2)+4F_{,\phi}^2/F}
{(1+3r_1)(f_{,X}+2r_2)+3F_{,\phi}^2/F}\,,\qquad
\frac{k^2}{a^2}\Phi \simeq
-\frac{\delta \rho}{2F} \frac{(1+2r_1)(f_{,X}+2r_2)+2F_{,\phi}^2/F}
{(1+3r_1)(f_{,X}+2r_2)+3F_{,\phi}^2/F}\,,\nonumber \\
& & G_{\rm eff} \simeq \frac{1}{8\pi F} 
\frac{1+4r_1+F_{,\phi}^2/(\mu F)}{1+3r_1}\,, \qquad
\zeta \simeq \frac{(1+4r_1)(f_{,X}+2r_2)+4F_{,\phi}^2/F}
{(1+2r_1)(f_{,X}+2r_2)+2F_{,\phi}^2/F}\,,
\end{eqnarray}
where $\mu=(1+3r_1)(f_{,X}+2r_2)+3F_{,\phi}^2/F$.
These correspond to the generalization of the results obtained in Ref.~\cite{Tsujikawa07}
with the mass term $M_{\phi}$ taken into account ($r_2 \neq 0$).

There are three different regimes depending on the values of $r_1$ and $r_2$.

\begin{itemize}
\item (i) $r_1 \ll 1$ and $r_2 \gg F_{,\phi}^2/F$
\vspace{0.1cm}

This corresponds to the regime in which both the scalaron and the field $\phi$
are sufficiently heavy. The evolution of cosmological perturbations mimics that of
GR.

\vspace{0.1cm}

\item (ii) $r_1 \ll 1$ and $r_2 \ll F_{,\phi}^2/F$
\vspace{0.1cm}

In this regime the scalaron is sufficiently heavy, whereas the field $\phi$
is light such that it gives rise to the modification of gravity. In fact we have 
\begin{equation}
\label{interevo}
\frac{k^2}{a^2}\Psi \simeq -\frac{\delta \rho}{2F} 
\frac{f_{,X}+4F_{,\phi}^2/F}{f_{,X}+3F_{,\phi}^2/F}\,,
\quad
\frac{k^2}{a^2}\Phi \simeq -\frac{\delta \rho}{2F} 
\frac{f_{,X}+2F_{,\phi}^2/F}{f_{,X}+3F_{,\phi}^2/F}\,,
\quad
G_{\rm eff} \simeq \frac{1}{8\pi F} 
\frac{f_{,X}+4F_{,\phi}^2/F}{f_{,X}+3F_{,\phi}^2/F}\,,\quad
\zeta=\frac{f_{,X}+4F_{,\phi}^2/F}{f_{,X}+2F_{,\phi}^2/F}\,.
\end{equation}
Let us consider Brans-Dicke (BD) theory \cite{BDicke} 
with the field potential, i.e.
\begin{equation}
f=\frac{\phi}{\kappa} R+\frac{2\omega_{\rm BD}}
{\kappa \phi}X-2V(\phi)\,,
\label{BDaction}
\end{equation}
where $\omega_{\rm BD}$ is the BD parameter, and 
$\kappa=1/M_{\rm pl}$ ($M_{\rm pl}$ is the 
reduced Planck mass). It then follows that 
\begin{equation}
\frac{k^2}{a^2}\Psi \simeq -\frac{\delta \rho}{2F} 
\frac{2\omega_{\rm BD}+4}{2\omega_{\rm BD}+3}\,,
\quad
\frac{k^2}{a^2}\Phi \simeq -\frac{\delta \rho}{2F} 
\frac{2\omega_{\rm BD}+2}{2\omega_{\rm BD}+3}\,,
\quad
G_{\rm eff} \simeq \frac{1}{8\pi F} 
\frac{2\omega_{\rm BD}+4}{2\omega_{\rm BD}+3}\,,\quad
\zeta=\frac{\omega_{\rm BD}+2}{\omega_{\rm BD}+1}\,.
\end{equation}
The no ghost condition in BD theory corresponds to $\omega_{\rm BD}>-3/2$, 
in which case the gravitational coupling $G_{\rm eff}$ is 
positive. The evolution of perturbations during the matter-dominated 
epoch is given by 
\begin{equation}
\label{delmsca}
\delta_m \propto t^{\frac16\left( \sqrt{\frac{99+50\omega_{\rm BD}}
{3+2\omega_{\rm BD}}}-1 \right)}\,,\qquad
\Phi_{\rm eff} \propto t^{\frac16\left( \sqrt{\frac{99+50\omega_{\rm BD}}
{3+2\omega_{\rm BD}}}-5 \right)}\,.
\end{equation}
Compared to GR ($\omega_{\rm BD} \to \infty$), the growth rate of 
perturbations is enhanced for the theories with $\omega_{\rm BD}>-3/2$.
We have $\zeta>1$ for $\omega_{\rm BD}>-1$ and 
$\zeta<-1$ for $-3/2<\omega_{\rm BD}<-1$.

\vspace{0.1cm}

\item (iii) $r_1 \gg 1$
\vspace{0.1cm}

This is the regime in which the scalaron has a light mass.
Irrespective of the values of $r_2$ one has
\begin{equation}
\frac{k^2}{a^2}\Psi \simeq -\frac{\delta \rho}{2F} 
\cdot \frac{4}{3}\,,
\qquad
\frac{k^2}{a^2}\Phi \simeq -\frac{\delta \rho}{2F} 
\cdot \frac{2}{3}\,,
\qquad
G_{\rm eff}=\frac{1}{8\pi F} \cdot \frac43\,,
\qquad
\zeta=2\,,
\end{equation}
which correspond to the massless regime $r_1 \gg 1$
in $f(R)$ gravity.
\end{itemize}

We can consider models in which the sequence of three 
cosmological epochs (i) $\to$ (ii) $\to$ (iii) occurs.
If this occurs during the matter era, the matter perturbation
starts to evolve as $\delta_m \propto t^{2/3}$ in the region (i) 
and its evolution is followed by (\ref{delmsca}) [region (ii)] 
and (\ref{delmfR}) [region (iii)].
In this case the variable $\zeta$ evolves as 1 $\to$ 
$(\omega_{\rm BD}+2)/(\omega_{\rm BD}+1)$ $\to$ 2. 
If the perturbations reach the regime $r_1>1$ before entering the regime 
$r_2<F_{,\phi}^2/F$, then there is no intermediate 
regime (ii) characterized by Eq.~(\ref{interevo}).

\subsection{Brans-Dicke theory with $\xi \neq 0$}

Let us consider BD theory with a potential described by the Lagrangian 
(\ref{BDaction}) in the presence of the field self-interaction 
$\xi (\phi) \square \phi (\partial^{\mu} \phi \partial_{\mu} \phi)$.
{}From Eqs.~(\ref{Psieqfi}), (\ref{Phieqfi}), (\ref{Geff}), (\ref{zeta}),
and (\ref{Phieff}), it follows that 
\begin{eqnarray}
& & \frac{k^2}{a^2}\Psi \simeq 
-\frac{\kappa \delta \rho}{2\phi} 
\frac{2\omega_{\rm BD}+4+2r_2 \kappa \phi+8\xi \kappa \phi 
(\ddot{\phi}+2H\dot{\phi})}{2\omega_{\rm BD}+3+2r_2 \kappa \phi+
8\xi \kappa \phi (\ddot{\phi}+2H\dot{\phi})-4\xi \kappa 
(\dot{\phi}^2+\xi \kappa \dot{\phi}^4)}\,, 
\label{BDxi1}
\\
& & \frac{k^2}{a^2}\Phi \simeq 
-\frac{\kappa \delta \rho}{2\phi} 
\frac{2\omega_{\rm BD}+2+2r_2 \kappa \phi+8\xi \kappa \phi 
(\ddot{\phi}+2H\dot{\phi})-4\xi \kappa \dot{\phi}^2}
{2\omega_{\rm BD}+3+2r_2 \kappa \phi+
8\xi \kappa\phi (\ddot{\phi}+2H\dot{\phi})
-4\xi \kappa (\dot{\phi}^2+\xi \kappa\dot{\phi}^4)}\,, 
\label{BDxi2}
\\
& & G_{\rm eff} \simeq \frac{\kappa}{8 \pi \phi} \left[ 
1+\frac{(1+2\xi \kappa \dot{\phi}^2)^2}{\kappa \phi \mu} \right]\,,
\label{BDxi3}
\\
& & \zeta \simeq \frac{2\omega_{\rm BD}+4+2r_2 \kappa \phi
+8 \xi \kappa \phi (\ddot{\phi}+2H \dot{\phi})}
{2\omega_{\rm BD}+2+ 2r_2 \kappa \phi
+8 \xi \kappa \phi (\ddot{\phi}+2H \dot{\phi})-4\xi \kappa \dot{\phi}^2}\,,
\label{BDxi4}
\\
& & \Phi_{\rm eff} \simeq -\frac{\kappa \rho}{2\phi} \frac{a^2}{k^2} \delta_m
\frac{2\omega_{\rm BD}+3+2r_2 \kappa \phi+8\xi \kappa\phi 
(\ddot{\phi}+2H \dot{\phi})-2\xi \kappa \dot{\phi}^2}
{2\omega_{\rm BD}+3+2r_2 \kappa \phi+
8\xi \kappa\phi (\ddot{\phi}+2H \dot{\phi})
-4\xi \kappa (\dot{\phi}^2+\xi \kappa \dot{\phi}^4)}\,,
\label{BDxi5}
\end{eqnarray}
where 
\begin{equation}
\kappa \phi \mu=2\omega_{\rm BD}+3+2r_2 \kappa \phi
+2\xi \kappa [4 \phi (\ddot{\phi}+2H \dot{\phi})
-2\dot{\phi}^2-2\xi \kappa \dot{\phi}^4 ]\,.
\label{kappaphidef}
\end{equation}
For the special case with $r_2=0$ and $\xi (\phi)=1/(M \phi^2)$ ($M$
is a constant having a dimension of mass), the effective gravitational
coupling (\ref{BDxi3}) agrees with the one derived in Ref.~\cite{Silva}.

In the presence of the field potential $V(\phi)$, one can recover the results 
in GR by taking the massive limit $r_2 \kappa \phi \to \infty$ in 
Eqs.~(\ref{BDxi1})-(\ref{BDxi5}).
There is another mechanism called the Vainshtein mechanism \cite{Vainshtein}
for the recovery of GR in the high-curvature
regime (i.e. during radiation and deep matter eras).
This is the case in which the term $8\xi \kappa \phi (\ddot{\phi}+2H \dot{\phi})$
is the dominant contributions in Eqs.~(\ref{BDxi1})-(\ref{BDxi5}).
In fact this situation arises for the function $\xi(\phi)$
responsible for the late-time cosmic acceleration.
For the choice $\xi (\phi)=1/(M \phi^2)$, provided $\omega_{\rm BD}<-2$,
there is a stable de Sitter solution 
with $\dot{H}=0$ and $x \equiv \dot{\phi}/(H\phi)=$\,constant
even in the absence of the field potential ($r_2=0$) \cite{Silva,DeFelice10}.

For the choice $\xi=1/(M\phi^2)$, one has 
$\xi \kappa \phi (\ddot{\phi}+2H\dot{\phi}) \gg 1$
and $x \equiv \dot{\phi}/(H\phi) \ll 1$ 
in the early cosmological epoch \cite{Silva}.
During the matter era, for example, we have  that 
$\xi \kappa H \phi \dot{\phi}=\kappa H^2 x/M \simeq 1/(6x) \gg 1$, 
where we have used $y \equiv \kappa H^2 x^2/M \simeq 1/6$ \cite{DeFelice10}.
In this regime the perturbation equations (\ref{BDxi1})-(\ref{BDxi5}) 
recover the results in GR.
At late times the deviation from GR arises, so that
the evolution of perturbations is subject to change.
As we will see later, the avoidance for the appearance of 
ghosts demands the condition $x>0$.
Under this condition the $\kappa \phi \mu$ term
defined in Eq.~(\ref{kappaphidef}) can remain to be positive \cite{Silva}. 
Hence the growth rate of matter perturbations gets
larger than that in GR, unlike the  DGP model \cite{DGPper}.
We also note that the presence of the nonlinear self-interaction gives rise to 
a nontrivial contribution to $\Phi_{\rm eff}$ because the last fraction 
in Eq.~(\ref{BDxi5}) is different from 1.

The presence of the field potential $V(\phi)$ (i.e. $r_2 \neq 0$) gives 
rise to the change for the evolution of perturbations.
One may consider theories in which 
the field is heavy ($r_2 \kappa \phi \gg 1$) in the early cosmological epoch, 
but the presence of the nonlinear self-interaction leads to the recovery of GR instead of
the chameleon mechanism based on a heavy scalar field.
More specifically this corresponds to the condition 
$\xi \kappa \phi (\ddot{\phi}+2H \dot{\phi}) \gg r_2 \kappa \phi \gg 1$.
Depending on the evolution of the terms  
$\xi \kappa \phi (\ddot{\phi}+2H \dot{\phi})$ and $r_2 \kappa \phi$, 
the perturbations evolve differently at late times.

For more general theories with the Lagrangian 
$f=F(\phi)R+2\omega (\phi)X-2V(\phi)$ (i.e. $r_1=0$)
the cosmological Vainshtein mechanism mentioned above can be 
also at work, provided that the $\xi (\ddot{\phi}+2H\dot{\phi})$ term 
is the dominant contributions in Eqs.~(\ref{Psieqfi}), (\ref{Phieqfi}), 
(\ref{Geff}), (\ref{zeta}), and (\ref{Phieff}).

\section{Conditions for the avoidance of ghosts and instabilities}
\label{ghostsec}

In this section, we study the full Lagrangian perturbed at second order
without using any approximation to derive conditions for the avoidance 
of ghosts and Laplacian instabilities. Let us consider the following
general action 
\begin{equation}
\label{eq:gl1}
 S = \int {\rm d}^4x \sqrt{-g}
\left[ \frac{1}{2}f(R,\phi,X)
+ \xi (\phi) \square \phi (\partial^{\mu} \phi 
\partial_{\mu} \phi)
+ P(Z) \right], 
\qquad Z= - \frac{1}{2}\partial^{\mu}\chi\partial_{\mu}\chi\,,
\end{equation}
where $P(Z)$ models the Lagrangian for the matter fluid $\chi$ 
with the following parameterized equation of state 
\begin{equation}
\rho = 2ZP'(Z) - P(Z),\quad  P = P(Z)\,.
\end{equation}
Here a prime represents a derivative with respect to $Z$. 
The sound speed squared of this fluid is \cite{kinf}
\begin{equation}
c_s^2 = \frac{P'(Z)}{2ZP''(Z)+P'(Z)}\,.
\end{equation}
As long as scalar and tensor perturbations are concerned, the above 
k-essence description of the matter field $\chi$ is equivalent to 
the description for the barotropic fluid.

For this action we generally have three propagating scalar degrees of
freedom, one from gravity, one from the field $\phi$, and one from the
matter fluid. The action (\ref{eq:gl1}) can be written as
\begin{equation}
\label{eq:gl2}
 S = \int {\rm d}^4x \sqrt{-g}
  \left[ 
   \frac{1}{2}(R-\lambda)F
   + \frac{1}{2}f(\lambda,\phi,X)
   + \xi (\phi) \square \phi (\partial^{\mu} \phi 
\partial_{\mu} \phi)
   + P(Z) \right]\,,
\end{equation}
where the equations of motion for the auxiliary fields lead to
\begin{equation}
\lambda=R\,,
\qquad
F=\frac{\partial f}{\partial\lambda}\,.
\label{FW}
\end{equation}
Variation of the action (\ref{eq:gl2}) leads to the same equations of 
motion as those derived by varying the action (\ref{eq:gl1}). Note that
$F$ in general depends not only on $\lambda=R$ 
but also on $\phi$ and $X$.
One example of theories with $F_{,X}\ne 0$ is 
$f=M_{\rm pl}^2R+\alpha R^2+X/(R^2/M_{\rm pl}^4)^m-V(\phi)$ with 
$m>3/2$~\cite{Mukohyama:2003nw}.

Formally, the auxiliary field $\lambda$ can be eliminated by solving its
equation of motion $F=\partial f/\partial\lambda$. 
In practice, this procedure can be done at each order in perturbative expansion. 
At the linear order we have 
\begin{equation}
 \delta \lambda = 
  \frac{\delta F-f_{,R\phi}\delta\phi-f_{,RX}\delta X}{f_{,RR}}\,,
  \label{eqn:deltalambda}
\end{equation}
provided that $f_{,RR}\ne 0$. Using this relation to eliminate 
$\delta\lambda$ at the level of the action only makes sense when 
$f_{,RR}\ne 0$. In subsection \ref{subsec:fRRne0}, we shall consider
theories with $f_{,RR}\ne 0$. 

The case with $f_{,RR}=0$ has measure zero in the space of
theories. Indeed, adding e.g., $\alpha R^2$ with an extremely small
$\alpha$ to $f$ would make $f_{,RR}$ non-vanishing. Therefore,
physically speaking, it is not really necessary to study theories
with strictly vanishing $f_{,RR}$. Nonetheless, in subsection
\ref{subsec:fRR0} we shall consider the case with $f_{,RR}=f_{,RX}=0$ as 
a simple example of theories with $f_{,RR}=0$.
The case with $f_{,RR}=0$ and $f_{,RX}\ne 0$ is not only physically
irrelevant (having measure zero in the space of theories) but also
technically complicated. Thus, we shall not address this case in the
present paper.

\subsection{Case (i): $f_{,RR}\ne 0$}
\label{subsec:fRRne0}

We consider the flat FLRW background plus scalar-type 
perturbations $\alpha$, $\beta$, ${\cal R}$, and $E$: 
\begin{equation}
{\rm d} s^2 = -(1+2\alpha) {\rm d}t^2 + 2a(t)^2\beta_{,i}
{\rm d}t {\rm d}x^i
+ a(t)^2\left[(1-2{\cal R})\delta_{ij} + 2E_{,ij} 
\right] {\rm d}x^i {\rm d}x^j\,.
\end{equation}
By using the Faddeev-Jackiw method~\cite{Faddeev:1988qp}, 
we can calculate the reduced quadratic action for gauge-invariant
perturbations (see also Refs.~\cite{Suyama}). 
Since the perturbed Hamiltonian and momentum constraints
are of the form 
\begin{eqnarray}
 0 & = & \frac{\partial{\cal H}^{(2)}}{\partial\alpha}
  = -H\Pi_{\cal R} + \dot{\phi}\Pi_{\delta\phi}
  + \dot{F}\Pi_{\delta F}
  + \dot{\chi}\Pi_{\delta\chi}
  + h_1({\cal R},\delta\phi,\delta F,\delta\chi,E)\,, \nonumber\\
 0 & = & \frac{\partial{\cal H}^{(2)}}{\partial\beta} = 
  \Pi_E 
  + h_2({\cal R},\delta\phi,\delta F,\delta\chi,E)\,, 
\end{eqnarray}
where ${\cal H}^{(2)}$ is the quadratic Hamiltonian density and 
$\Pi_{{\cal R},\delta\phi,\delta F,\delta\chi,E}$ are momenta conjugate
to ${\cal R},\delta\phi,\delta F,\delta\chi,E$ respectively, we can solve
them with respect to $\Pi_{\delta\chi}$ and $\Pi_E$. 

Noting that the constraints are generators of gauge transformation, the form of the
constraints enables us to identify the three gauge-invariant variables as
\begin{equation}
 {\cal Q}_1 \equiv  
  \frac{{\cal R}}{H} + \frac{\delta\chi}{\dot{\chi}}\,, 
  \qquad
 {\cal Q}_2 \equiv  \frac{\delta\phi}{\dot{\phi}}
  - \frac{\delta\chi}{\dot{\chi}}\,, 
  \qquad
 {\cal Q}_3 \equiv  \frac{\delta F}{\dot{F}}
  - \frac{\delta\chi}{\dot{\chi}}\,.
\end{equation}
By substituting the solutions of the Hamiltonian and momentum
constraints to the quadratic action
\begin{equation}
S^{(2)} = \int {\rm d}^4x
\left[ \Pi_{\cal R}\dot{\cal R}
+\Pi_{\delta\phi}\delta\dot{\phi}
+\Pi_{\delta F}\delta\dot{F}
+\Pi_{\delta\chi}\delta\dot{\chi}
+\Pi_E\delta\dot{E}
-{\cal H}^{(2)} \right]\,,
 \end{equation} 
we obtain the reduced quadratic action written in terms of the 
gauge-invariant variables ${\cal Q}_i$ ($i=1,2,3$) and their conjugate momenta
$\Pi_{{\cal Q}_i}$ only. After using the equations of motion for the
conjugate momenta to express them in terms of ${\cal Q}_i$ and
$\dot{\cal Q}_i$, the reduced quadratic action is expressed in terms of
${\cal Q}_i$ and $\dot{\cal Q}_i$ as 
\begin{equation}
S^{(2)} = \frac{1}{2}\int {\rm d}^4x\,a^3
\left[ \dot{\vec{{\cal Q}}}^t {\bm K}\dot{\vec{{\cal Q}}}
-\frac{1}{a^2}
\nabla\vec{{\cal Q}}^t {\bm G}\nabla{\vec{{\cal Q}}}
-\vec{{\cal Q}}^t {\bm B}\dot{\vec{{\cal Q}}}
-\vec{{\cal Q}}^t {\bm M}\vec{{\cal Q}}
\right]\,, \label{eqn:reduced-action-Q}
\end{equation} 
where ${\bm K}$, ${\bm G}$, ${\bm B}$ and ${\bm M}$ are 
$3\times 3$ matrices. 

\subsubsection{No ghost conditions}

The three eigenvalues $K_i$ ($i=1,2,3$) of the kinetic term matrix 
${\bm K}$ are all positive if and only if the following three
combinations are positive. 
\begin{eqnarray}
 K_1K_2K_3 & = &  \alpha_1B/D\,, \nonumber\\
 K_1K_2+K_2K_3+K_3K_1 & = &  (\alpha_2B+\alpha_3)/D\,, \nonumber\\
 K_1+K_2+K_3 & = &  (\alpha_4B+\alpha_5)/D\,,
\end{eqnarray}
where
\begin{eqnarray}
 B & = &
  24\xi H\dot{\phi}-8\xi_{,\phi}\dot{\phi}^2
  +f_{,X}+f_{,XX}\dot{\phi}^2
  - F_{,X}^2\dot{\phi}^2/F_{,R}\,, 
  \nonumber\\
 D & = & (\dot{F} + 2HF -2\xi\dot{\phi}^3)^2\,,
  \nonumber\\
 \alpha_1 & = & 3(\rho+P)c_s^{-2}H^2\dot{\phi}^2\dot{F}^2F\,, 
  \nonumber\\
 \alpha_2 & = & 
  [(\dot{F}^2+4H^2F^2)(\rho+P)c_s^{-2}
   +18H^2\dot{F}^2F ]\dot{\phi}^2/2\,,
  \nonumber\\
 \alpha_3 & = & 6(\dot{F}^2+4\xi^2\dot{\phi}^6)(\rho+P)c_s^{-2}H^2F\,,
  \nonumber\\
 \alpha_4 & = & 
  [(\dot{F}+HF)^2+3H^2F^2]\dot{\phi}^2\,,
  \nonumber\\
 \alpha_5 & = & 
  (\dot{F}^2+4H^2F^2+4\xi^2\dot{\phi}^6)(\rho+P)c_s^{-2}
   + 12[(\dot{F}-\xi\dot\phi^3)^2+3\xi^2\dot{\phi}^6]H^2F\,,
\end{eqnarray}
and $F=f_{,R}$. 

We assume that $F>0$ since this is required for the absence of tensor ghosts.
Let us also suppose that $(\rho+P)c_s^{-2}=2(2ZP''+P')Z>0$, 
which is required for the absence of ghosts in the matter sector. 
Under these assumptions, $D$ and $\alpha_i$ ($i=1,\cdots,5$) are 
non-negative, and there are no ghosts 
in the scalar sector if and only if $B>0$, i.e.
\begin{equation}
24\xi H\dot{\phi}-8\xi_{,\phi}\dot{\phi}^2
+f_{,X}+f_{,XX}\dot{\phi}^2
- F_{,X}^2\dot{\phi}^2/F_{,R}> 0\,.
\label{noghost1}
\end{equation}
\subsubsection{Avoidance of Laplacian instabilities}

For large $k$ the dispersion relation is specified by 
\begin{equation}
 \det (\omega^2{\bm K}-k^2{\bm G}/a^2) = 0\,.
\end{equation}
This is equivalent to 
\begin{equation}
 \det (c^2{\bm K}-{\bm G}) = 0\,,
\end{equation}
where $c$ is the speed of propagation defined by
\begin{equation}
\omega^2 = c^2\frac{k^2}{a^2}\,. 
\end{equation}

More explicitly we have 
\begin{equation}
 \det (c^2{\bm K}-{\bm G})
  = 3(\rho+P)BH^2\dot{\phi}^2\dot{F}^2F
  (c^2-c_s^2)
  \left[c^4 - (c_1^2+c_2^2)c^2 + c_1^2c_2^2\right]/D\,,
\end{equation}
where
\begin{equation}
 c_1^2+c_2^2 = 1 + \alpha_6/B\,,\qquad
 c_1^2c_2^2  = (\alpha_6-\alpha_7)/B\,,
\end{equation}
and
\begin{equation}
 \alpha_6 = f_{,X} + 8(\ddot{\phi}+2H\dot{\phi})\xi\,,\qquad
 \alpha_7 = 16\xi^2\dot{\phi}^4/(3F)\,. 
\end{equation}
Gradient instabilities are avoided if and only if $c_s^2$,
$c_1^2+c_2^2$, $c_1^2c_2^2$ and $(c_1^2-c_2^2)^2$ 
are all non-negative, where
\begin{equation}
 (c_1^2-c_2^2)^2 
  = \left(1-\alpha_6/B \right)^2
  +4\alpha_7/B\,.
\end{equation}

Under the conditions $F>0$ (absence of tensor ghosts) and $B>0$ (absence
of scalar ghosts), the condition for the absence of gradient
instabilities is equivalent to $\alpha_6>\alpha_7$, i.e.
\begin{equation}
f_{,X} + 8(\ddot{\phi}+2H\dot{\phi})\xi-16\xi^2\dot{\phi}^4/(3F)>0\,.
\label{noins1}
\end{equation}
\subsubsection{Mass of propagating modes}

We provide here a straightforward procedure on how to define the mass
for the propagating modes. For the reduced action
(\ref{eqn:reduced-action-Q}), we can first diagonalize and normalize the
kinetic matrix ${\bm K}$ to the identity matrix, through a convenient
linear field redefinition 
$\vec {\cal Q}={\bm A}\,\vec{\cal Q}'$. Each field redefinition, in the $k\to0$ limit, 
will contribute to the matrices ${\bm B}$ and 
${\bm M}$, as ${\bm A}$ is time-dependent and the modes feel the time
evolution of the background. 
Having the canonically normalized kinetic matrix, we can perform 
a rotation $\vec {\cal Q}'={\bm R}\,\vec {\cal Q}''$,
which transforms the mass matrix into 
\begin{equation}
{\bm M}''={\bm M}'+{\bm T}^2
-({\bm B}' {\bm T} + {\bm T} {\bm B}')/2\,,
\end{equation}
where ${\bm T}=\dot{{\bm R}}^t{\bm R}$ is antisymmetric. 
Then the mass eigenvalues are obtained by choosing the elements of 
${\bm T}$ which diagonalize ${\bm M}''$. This procedure implies solving 
a differential equation for the elements of ${\bm T}$ and $\dot{{\bm T}}$. 
We shall not present explicit expressions as they are rather complicated.

\subsubsection{Specific theories}

Let us apply the above results to the relaxation mechanism of
the cosmological constant~\cite{Mukohyama:2003nw}. 
In the simplest case the Lagrangian is specified as
\begin{equation}
f = M_{\rm pl}^2R + \alpha R^2
+ \frac{X}{(R^2/M_{\rm pl}^4)^m}-V(\phi),
\qquad \xi=0,
\end{equation}
with $m>3/2$. 
It is straightforward to show that
\begin{equation}
B=f_{,X} - \frac{F_{,X}^2\dot{\phi}^2}{F_{,R}}
=\frac{1}{(R^2/M_{\rm pl}^4)^m}
\left[1 - \frac{4m^2(R^2/M_{\rm pl}^4)^{m-1}}{F_{,R}}
\frac{\pi^2}{M_{\rm pl}^4}\right],
\end{equation}
where
\begin{equation}
\pi = \frac{\dot{\phi}}{(R^2/M_{\rm pl}^4)^m}\,,
\end{equation}
is the momentum conjugate to $\phi$ and 
\begin{equation}
F_{,R} = 2\alpha 
+ m(2m+1)(R^2/M_{\rm pl}^4)^{m-1}
\frac{\pi^2}{M_{\rm pl}^4}\,.
\end{equation}

The attractor behavior
\begin{equation}
\pi^2 \sim \left(\frac{M_{\rm pl}^3}{H}\right)^2\,,
\end{equation}
and $R\sim H^2$ imply that
\begin{equation}
(R^2/M_{\rm pl}^4)^{m-1}\frac{\pi^2}{M_{\rm pl}^4}
\sim (H/M_{\rm pl})^{2(2m-3)} \to 0 \qquad ({\rm as}~~H\to 0)\,,
\end{equation}
under the assumption $m>3/2$. 
Therefore, at low energy, we have
\begin{equation}
F_{,R} \sim 2\alpha, \qquad 
B\sim \frac{1}{(R^2/M_{\rm pl}^4)^m} > 0\,.
\end{equation}
This shows the absense of ghost. 
Gradient instability is also absent as
\begin{equation}
\alpha_6 = f_{,X} = \frac{1}{(R^2/M_{\rm pl}^4)^m} > 0\,, 
\qquad \alpha_7 = 0\,.
\end{equation}
Indeed, the two speeds of propagation $c_1$ and $c_2$ 
are both unity in the low-energy regime.

\subsection{Case (ii): $f_{,RR}=f_{,RX}=0$}
\label{subsec:fRR0}

In this subsection we consider the case with $f_{,RR}=f_{,RX}=0$ as a
simple example of theories with $f_{,RR}=0$. 
As already stated, theories with $f_{,RR}=0$ have measure zero in the space
of theories in the sense that they can easily be transformed to theories
with $f_{,RR}\ne 0$ by inclusion of an additional term like $\alpha R^2$
with small $\alpha$ to $f$. 

The theories with $f_{,RR}=f_{,RX}=0$ correspond to
$f=F(\phi)R+f_2(X,\phi)$. One example of the theories with
$f_{,RR}=f_{,RX}=0$ and $f_{,XX} \neq 0$ is
$f=F(\phi)R-2X+2X^2/M^4$. Since 
$\delta F=F_{,\phi}\,\delta\phi$, the two fields ${\cal Q}_1$ and 
${\cal Q}_2$ are sufficient to study the perturbed action. Although this
fact simplifies the problem, the nonminimal coupling to $R$ directly
affects the propagation properties of the field $\delta\phi$. 

Since $\delta F=F_{,\phi}\delta\phi$, we consider the original action
(\ref{eq:gl1}) instead of the equivalent action
(\ref{eq:gl2}). Specializing to the present case, we have
\begin{equation}
S = \int {\rm d}^4x \sqrt{-g}
\left[ \frac{1}{2}F(\phi)R + \frac{1}{2}f_2(\phi, X)
+ \xi (\phi) \square \phi (\partial^{\mu} \phi 
\partial_{\mu} \phi)
+ P(Z) \right]\,.
\end{equation}
By using the Faddeev-Jackiw method~\cite{Faddeev:1988qp}, we can
calculate the reduced quadratic action for two gauge-invariant
variables. The perturbed Hamiltonian and momentum constraints
are of the form 
\begin{eqnarray}
 0 & = & \frac{\partial{\cal H}^{(2)}}{\partial\alpha}
  = -H\Pi_{\cal R} + \dot{\phi}\Pi_{\delta\phi}
  + \dot{\chi}\Pi_{\delta\chi}
  + h_1({\cal R},\delta\phi,\delta\chi,E)\,, \nonumber\\
 0 & = & \frac{\partial{\cal H}^{(2)}}{\partial\beta} = 
  \Pi_E 
  + h_2({\cal R},\delta\phi,\delta\chi,E)\,. 
\end{eqnarray}

As in the previous case, we can solve them with respect to
$\Pi_{\delta\chi}$ and $\Pi_E$. The form of the constraints enables us
to identify the two gauge-invariant variables as 
\begin{equation}
 {\cal Q}_1 \equiv  
  \frac{{\cal R}}{H} + \frac{\delta\chi}{\dot{\chi}}\,, 
 \qquad
 {\cal Q}_2 \equiv  \frac{\delta\phi}{\dot{\phi}}
  - \frac{\delta\chi}{\dot{\chi}}\,. 
\end{equation}
By substituting the solutions of the Hamiltonian and momentum
constraints to the quadratic action and expressing the conjugate momenta
in terms of ${\cal Q}_i$ and $\dot{\cal Q}_i$, we obtain the reduced
quadratic action of the form, 
\begin{equation}
S^{(2)} = \frac{1}{2}\int {\rm d}^4x\,a^3
\left[ \dot{\vec{{\cal Q}}}^t {\bm K}\dot{\vec{{\cal Q}}}
-\frac{1}{a^2}\nabla\vec{{\cal Q}}^t {\bm G}\nabla{\vec{{\cal Q}}}
-\vec{{\cal Q}}^t {\bm B}\dot{\vec{{\cal Q}}}
-\vec{{\cal Q}}^t {\bm M}\vec{{\cal Q}} \right]\,,
\end{equation} 
where ${\bm K}$, ${\bm G}$, ${\bm B}$ and ${\bm M}$ are 
now $2\times 2$ matrices.

\subsubsection{No ghost conditions}

The two eigenvalues $K_i$ ($i=1,2$) of the kinetic term matrix 
${\bm K}$ are all positive if and only if the following two
combinations are positive:
\begin{eqnarray}
 K_1K_2 & = &  \tilde{\alpha}_1\tilde{B}/\tilde{D}\,, \nonumber\\
 K_1+K_2 & = &  (\tilde{\alpha}_2\tilde{B}+\tilde{\alpha}_3)/\tilde{D}\,,
\end{eqnarray}
where
\begin{eqnarray}
 \tilde{B} & = &
  (24\xi H\dot{\phi}-8\xi_{,\phi}\dot{\phi}^2
  +f_{,X}+f_{,XX}\dot{\phi}^2 )F\dot{\phi}^2
  + 3(\dot{F}-2\xi\dot{\phi}^3)^2\,,
  \nonumber\\
 \tilde{D} & = & (\dot{F} + 2HF -2\xi\dot{\phi}^3)^2\,,
  \nonumber\\
 \tilde{\alpha}_1 & = & 2(\rho+P)c_s^{-2}H^2F\,, 
  \nonumber\\
 \tilde{\alpha}_2 & = & 4H^2F
  \nonumber\\
 \tilde{\alpha}_3 & = & 
  [4F^2H^2+(\dot{F}-2\xi\dot{\phi}^3)^2 ]
  (\rho+P)c_s^{-2}\,.
\end{eqnarray}

The absence of ghosts for tensor perturbations demands the condition 
$F>0$. The absence of ghosts in the matter sector requires that 
$(\rho+P)c_s^{-2}=2(2ZP''+P')Z>0$.
Under these assumptions, $\tilde{D}$ and 
$\tilde{\alpha}_i$ ($i=1,2,3$) are non-negative, and 
there are no ghosts in the scalar sector 
if and only if $\tilde{B}>0$, i.e. 
\begin{equation}
(24\xi H\dot{\phi}-8\xi_{,\phi}\dot{\phi}^2
+f_{,X}+f_{,XX}\dot{\phi}^2 )F\dot{\phi}^2
+ 3(\dot{F}-2\xi\dot{\phi}^3)^2>0\,.
\label{abghost}
\end{equation}
\subsubsection{Avoidance of Laplacian instabilities}

For large $k$ the dispersion relation is given by 
\begin{equation}
 \det (c^2{\bm K}-{\bm G}) = 0\,,
\end{equation}
where $c$ is the speed of propagation defined by
$\omega^2 = c^2\,k^2/a^2$.

There are two solutions to this equation: one is $c^2=c_s^2$ 
and another is
\begin{equation}
 c^2 = \frac{[f_{2,X}+8\xi(\ddot{\phi}+2H\dot{\phi})]
  F\dot{\phi}^2
  +(3\dot{F}+2\xi\dot{\phi}^3)(\dot{F}-2\xi\dot{\phi}^3)}
  {(24\xi H\dot{\phi}-8\xi_{,\phi}\dot{\phi}^2
  +f_{,X}+f_{,XX}\dot{\phi}^2 )F\dot{\phi}^2
  + 3(\dot{F}-2\xi\dot{\phi}^3)^2}\,.
  \label{cap}
\end{equation}
Under the no ghost condition (\ref{abghost}), the
condition for the absence of gradient instabilities 
corresponds to 
\begin{equation}
[f_{2,X}+8\xi(\ddot{\phi}+2H\dot{\phi})]
F\dot{\phi}^2+(3\dot{F}+2\xi\dot{\phi}^3)
(\dot{F}-2\xi\dot{\phi}^3)>0\,.
\label{noins2}
\end{equation}
\subsubsection{Specific theories}

Let us apply the above results to generalized Brans-Dicke theories
recently proposed in Ref.~\cite{DeFelice10}.
These theories are given by 
\begin{equation}
F(\phi)=\kappa^{1-n}\phi^{3-n}\,,
\qquad
f_2(\phi, X)=2\omega (\kappa \phi)^{1-n}X\,,
\qquad
\xi(\phi)=M^{n-3}\phi^{-n}\,,
\label{spefunctions}
\end{equation}
which can be obtained by demanding the existence of 
de Sitter solutions responsible for dark energy.
The Brans-Dicke theory with $\xi (\phi)=1/(M\phi^2)$ \cite{Silva}
corresponds to the choice $n=2$. 
The viable model parameter space is restricted in the 
regions $2 \le n \le 3$ and $\omega<-n(n-3)^2$ \cite{DeFelice10}.

For this theory the no ghost condition (\ref{abghost}) reduces to 
\begin{equation}
\tilde{B}=(FH)^2 x \left[ 24 y+x \left\{ 2\omega+8ny
+3(3-n-2y)^2 \right\} \right]>0\,,
\label{tilBcon}
\end{equation}
where $x \equiv \dot{\phi}/(H\phi)$ and 
$y \equiv x^2M^{n-3}H^2/\kappa^{1-n}>0$.
During the radiation and deep matter eras 
one has $x \ll y$ for the theories with $2 \le n<3$.
Since $\tilde{B} \simeq 24(FH)^2xy$ in this regime, 
we require that $x>0$ to avoid the appearance of ghosts.
The sign change of $x$ leads to the violation of the condition 
(\ref{tilBcon}) and hence the condition $x>0$ 
is required during the cosmological evolution 
from the radiation era to the de Sitter epoch \cite{DeFelice10}.

Let us consider the evolution of the propagation speed (\ref{cap})
during the radiation and matter eras for $2 \le n <3$.
In this regime the dominant terms in the numerator and 
the denominator of Eq.~(\ref{cap}) correspond to
$8 \xi (\ddot{\phi}+2H\dot{\phi})F \dot{\phi}^2$ and
$24\xi H F\dot{\phi}^3$, respectively. This gives
\begin{equation}
c^2 \simeq \frac23+\frac{\ddot{\phi}}{3H \dot{\phi}}\,.
\label{c2}
\end{equation}
{}From Eqs.~(\ref{be2}) and (\ref{be3}) we also obtain 
the following approximate relation 
\begin{equation}
4 H \xi \dot{\phi} \ddot{\phi}+2\xi (3H^2+\dot{H}) \dot{\phi}^2
-F_{,\phi} (2H^2+\dot{H}) \simeq 0\,.
\label{Hxi}
\end{equation}
As long as the field energy density is suppressed relative 
to the radiation and matter densities we have 
$\dot{H}/H^2 \simeq -2+\Omega_m/2$, where 
$\Omega_m \equiv \rho_m/(3FH^2)$ is the 
density parameter of matter.
Plugging Eq.~(\ref{Hxi}) into Eq.~(\ref{c2}), 
it follows that 
\begin{equation}
c^2 \simeq \frac{1}{12} (6-\Omega_m)
+\frac{3-n}{24y}\Omega_m\,.
\end{equation}
Since $y \simeq (3-n)\Omega_m/8$ during the radiation 
era \cite{DeFelice10}, one has $c^2 \simeq 5/6-\Omega_m/12$. 
Meanwhile $y \simeq (3-n)/6$ during the matter era 
($\Omega_m \simeq 1$), so that $c^2 \simeq 2/3$. 
These results agree with those obtained in 
Refs.~\cite{Silva,DeFelice10}. 
At the de Sitter point one has $0 \le c^2<1$ as long as
$x>0$ and $n \ge 2$ \cite{DeFelice10}.

\section{Conclusions}

We have discussed cosmological perturbations for the general 
Lagrangian $f (R,\phi, X)/2+\xi (\phi) \square \phi 
(\partial^{\mu} \phi \partial_{\mu} \phi)$, which includes 
most of modified gravity theories proposed in literature.
The presence of the nonlinear field self-interaction term allows 
a possibility for the recovery of GR in the early cosmological epoch, 
even if the field potential is absent.
We also take into account the effect of the field potential 
to cover the case in which the transition from the ``massive'' (GR) 
regime to the ``massless'' (scalar-tensor) regime occurs
during the cosmological evolution by today.

Under the quasi-static approximation on sub-horizon scales
we have derived the equations for the matter perturbation $\delta_m$ 
and gravitational potentials $\Psi$ and $\Phi$ for the theories
$f=f_1(R, \phi)+f_2 (\phi, X)$, see Eqs.~(\ref{Psieqfi})-(\ref{muexp}).
There are two important mass scales given by 
$M_R^2=f_{,R}/(3f_{,RR})$ and $M_\phi^2=-f_{,\phi \phi}/2$, 
which come from the gravitational scalar degree of freedom and
the field $\phi$ respectively.
As long as both scalars are massive such that 
$\{ M_R^2, M_{\phi}^2 \} \gg k^2/a^2$, the perturbations 
are in the GR regime characterized by 
$\delta_m \propto t^{2/3}$ and $\Phi_{\rm eff}=(\Psi+\Phi)/2=$\,constant
during the matter-dominated epoch.
The evolution of perturbations is subject to change in the regime 
$M_R^2 \lesssim k^2/a^2$ or $M_{\phi}^2 \lesssim k^2/a^2$.
The epoch at which the transition to the massless regime occurs 
depends on the models of dark energy. 

In Sec.~\ref{specific} we have applied the results derived under the quasi-static 
approximation to a number of modified gravity theories and 
estimated the growth rate of perturbations.
Not only with massive fields satisfying the conditions 
$\{ M_R^2, M_{\phi}^2 \} \gg k^2/a^2$ but also 
in the presence of the nonlinear field self-interaction, 
the evolution of perturbations mimics that 
in GR during the early cosmological evolution.
We showed that our results recover those obtained in Brans-Dicke theory 
with $V(\phi)=0$ and $\xi(\phi)=1/(M \phi^2)$ \cite{Silva,Kobayashi} 
as a special case.

In Sec.~\ref{ghostsec} we have derived the conditions for the avoidance 
of ghosts and Laplacian instabilities by deriving the full second-order action 
without using any approximation.
We have discussed two classes of theories: (i) $f_{,RR} \neq 0$
and (ii) $f_{,RR}=f_{,RX}=0$.
The case (i) corresponds to theories with three scalar-field
degrees of freedom, whereas in the case (ii) two scalar-field
degrees of freedom are present.
In the case (i) the conditions for the avoidance of ghosts and 
Laplacian instabilities are given by Eqs.~(\ref{noghost1}) 
and (\ref{noins1}) respectively, whereas in the case (ii) they are given by  
Eqs.~(\ref{abghost}) and (\ref{noins2}) respectively. 
We have also shown that in generalized Brans-Dicke 
theories described by the functions (\ref{spefunctions})
our results reproduce the no ghost condition as well as the propagation 
speed derived in Ref.~\cite{DeFelice10}.

It will be of interest to apply our results to the construction of 
viable modified gravity models of dark energy.
In particular the genericity of the cosmological Vainshtein mechanism 
due to the scalar-field self-interaction should be explored further, 
together with local gravity constraints.
The evolution of matter perturbations as well as gravitational 
potentials can be used to place constraints on modified gravity 
models from the observations of large-scale structure, 
CMB, and weak lensing. 
We hope that future observations and experiments will provide 
some distinguished features for the modification of gravity from 
General Relativity.

\section*{ACKNOWLEDGEMENTS}

The work of A.\,D and S.\,T. was supported by the Grant-in-Aid for Scientific 
Research Fund of the JSPS Nos.~09314 and 30318802. 
S.\,T. also thanks financial support for the Grant-in-Aid for 
Scientific Research on Innovative Areas (No.~21111006). 
The work of S.M. is supported by JSPS Grant-in-Aid for Young Scientists
(B) No.~17740134, JSPS Grant-in-Aid for Creative Scientific Research
No.~19GS0219, MEXT Grant-in-Aid for Scientific Research on Innovative
Areas No.~21111006, JSPS Grant-in-Aid for Scientific Research (C)
No.~21540278, the Mitsubishi Foundation, and World Premier International 
Research Center Initiative.


\end{document}